\PassOptionsToPackage{dvipsnames}{xcolor}
\documentclass[10pt,sigconf,nonacm,screen]{acmart}

\usepackage[all]{nowidow}
\usepackage{xcolor}
\usepackage{subcaption}
\usepackage{hyperref}
\usepackage{acronym}
\usepackage{csquotes}

\usepackage[capitalise,noabbrev]{cleveref}
\crefformat{section}{\S#2#1#3} 
\crefformat{subsection}{\S#2#1#3}
\crefformat{subsubsection}{\S#2#1#3}
\crefrangeformat{section}{\S#3#1#4 to~\S#5#2#6}
\crefrangeformat{subsection}{\S#3#1#4 to~\S#5#2#6}
\crefrangeformat{subsubsection}{\S#3#1#4 to~\S#5#2#6}

\usepackage{listings}
\begin{document}

\author{Natalie Carl}
\affiliation{%
    \institution{TU Berlin}
    \city{Berlin}
    \country{Germany}}
\email{nc@3s.tu-berlin.de}
\orcid{0009-0000-5991-9255}

\author{Trever Schirmer}
\affiliation{%
    \institution{TU Berlin}
    \city{Berlin}
    \country{Germany}}
\email{ts@3s.tu-berlin.de}
\orcid{0000-0001-9277-3032}

\author{Niklas Kowallik}
\affiliation{%
    \institution{TU Berlin}
    \city{Berlin}
    \country{Germany}}
\email{nk@3s.tu-berlin.de}
\orcid{0009-0006-9839-1864}

\author{Joshua Adamek}
\affiliation{%
    \institution{TU Dortmund}
    \city{Dortmund}
    \country{Germany}}
\email{joshua.adamek@tu-dortmund.de}
\orcid{0009-0004-5948-6903}

\author{Tobias Pfandzelter}
\affiliation{%
    \institution{TU Berlin}
    \city{Dortmund}
    \country{Germany}}
\email{tp@3s.tu-berlin.de}
\orcid{0000-0002-7868-8613}

\author{Sergio Lucia}
\affiliation{%
    \institution{TU Dortmund}
    \city{Dortmund}
    \country{Germany}}
\email{sergio.lucia@tu-dortmund.de}
\orcid{0000-0002-3347-5593}

\author{David Bermbach}
\affiliation{%
    \institution{TU Berlin}
    \city{Berlin}
    \country{Germany}}
\email{db@3s.tu-berlin.de}
\orcid{0000-0002-7524-3256}

\title{Multi-Event Triggers for Serverless Computing} 

\keywords{function-as-a-service, function triggers, serverless workflows}

\copyrightyear{2023}
\acmYear{2023}
\setcopyright{acmcopyright}\acmConference[Conference '24]{29th International Conference}{December 26--31, 2024}{Pyongyang, People's Republic of Korea} 
\acmBooktitle{29th International Conference (Conference '24), December 26--31, 2024, Pyongyang, People's Republic of Korea}

\begin{abstract}
    Function-as-a-Service (FaaS) is an event-driven serverless cloud computing model in which small, stateless functions are invoked in response to events, such as HTTP requests, new database entries, or messages.
    Current FaaS platform assume that each function invocation corresponds to a single event.
    However, from an application perspective, it is desirable to invoke functions in response to a collection of events of different types or only with every n\textsuperscript{th} event.
    To implement this today, a function would need additional state management, e.g., in a database, and custom logic to determine whether its trigger condition is fulfilled and the actual application code should run.
    In such an implementation, most function invocations would be rendered essentially useless, leading to unnecessarily high resource usage, latency, and cost for applications.
    In this paper, we introduce multi-event triggers, through which complex conditions for function invocations can be specified.
    Specifically, we introduce abstractions for invoking functions based on a set of $n$ events and joins of multiple events of different types.
    This enables application developers to define intricate conditions for function invocations, workflow steps, and complex event processing.
    Our evaluation with a proof-of-concept prototype shows that this reduces event--invocation latency by 62.5\% in an incident detection use-case and that our system can handle more than 300,000 requests per second on limited hardware, which is sufficient load for implementation in large FaaS platforms.
\end{abstract}

\maketitle

\section{Introduction}\label{sec:introduction}

FaaS is a serverless cloud computing model that enables developers to focus on implementing their business logic in small, stateless functions, while the platform handles scaling and manages the execution environment~\cite{brooker2023demand}.
Such functions are triggered by events, which can be HTTP requests, pub/sub messages, changes in databases, or events created periodically at certain times of day, among others.
FaaS follows a pay-per-use model, in which providers only charge developers for the resources their application actually consumes.

FaaS has been applied to different contexts and at different scales, ranging from traditional cloud platforms in large data centers, e.g., Amazon Web Services (AWS) Lambda or Google Cloud Platform (GCP) Functions, to edge settings in which FaaS functions are deployed to smaller devices, such as Raspberry Pis and microcontrollers~\cite{george_nanolambda_2020,pfandzelter_tinyfaas_2020}.
FaaS can serve as the central level of abstraction for building entire applications with multiple functions chained together to create serverless workflows~\cite{eismann_review_2021}.
Applications of serverless computing include data processing pipelines, machine learning, video processing, scientific computing, among others~\cite{eismann_review_2021,carreira_cirrus_2019,fouladi_encoding_nodate,chard_funcx_2020,muller_lambada_2020,pu_shuffling_nodate,jiang_towards_2021}.

Implementing such complex applications requires functions to invoke other functions, e.g., subsequent workflow steps.
This is trivial when a function invokes a single or multiple other functions.
However, invoking a single function with the results of multiple functions or other services-in essence a join or merge operation-cannot be built using today's abstractions.
We refer to the concept of triggering a single function with multiple events as \emph{multi-event triggers}.

To implement multi-event triggers today, a function has to keep state and, hence, needs access to a database for storing individual events and manually determining whether its trigger condition is fulfilled and the actual function logic should be executed.
Depending on the condition's complexity, the amount of times a function is executed to only update its event database can exceed the amount of times it executes its application logic by orders of magnitude.
Due to pay-per-use billing, these additional function invocations result in high costs for application developers.
This is exacerbated with expensive function cold-starts when events arrive in irregular intervals.
Most importantly, handling events manually goes against the FaaS model, where the platform itself is supposed to handle the arrival and distribution of events as well as the eventual invocation of a function~\cite{baldini_serverless_2017}.

In this paper, we attempt to close this gap by extending the event-based trigger mechanism in serverless computing to support multi-event triggers.
We aim to retain the benefits of the FaaS programming model, e.g., loose coupling, automatic scalability, and ease of development, by integrating multi-event trigger logic into the FaaS platform's function trigger mechanism.
Specifically, we make the following contributions:

\begin{itemize}
    \item We propose multi-event triggers as a novel extension of the FaaS function trigger mechanism, along with types of multi-event trigger rules useful to build FaaS applications (\cref{sec:arch1}).
    \item We introduce the design of a multi-event trigger engine to process events and trigger rules for FaaS platforms (\cref{sec:arch2}).
    \item We develop a prototype implementation of our multi-event trigger engine to demonstrate the feasibility of our design (\cref{sec:implementation}).
    \item We evaluate the potential of multi-event triggers in an incident detection use case (\cref{sec:eval}).
\end{itemize}

\section{FaaS Trigger Mechanisms}\label{sec:background}

Most FaaS platforms offer request-response patterns, pub/sub-subscripions, or messages queues for function invocation, as well as custom---usually platform-specific---solutions for triggering FaaS functions.
Virtually all FaaS platforms support synchronous triggers that use a request-response pattern, such as a RESTful interface through HTTP(S)~\cite{shahrad_serverless_2020,eismann_state_2022}.
Of course, functions can also use these direct communication mechanisms to trigger other functions to build larger workflows.

Similarly, functions can be invoked asynchronously through message queues or pub/sub systems.
AWS's event source mapping allows functions to process collections of events in a single invocation, resulting in a model adjacent to stream-processing.

Lastly, serverless functions can be invoked in response to events originating from other cloud services.
AWS and GCP, for example, allow the connection between FaaS functions and a variety of events their respective other services, ranging from changes to databases, at specific times, log messages, or infrastructure changes to invoking functions using email.
This wide collection of different trigger mechanisms allows FaaS to be used in a large variety of contexts—as glue code for responding to or handling infrastructure changes, for processing streams of data, or for building applications in distributed and heterogeneous edge environments.

The mechanics of trigger mechanisms are essential for serverless workflow execution, as they determine when a workflow is invoked and how functions communicate.
While $1:1$ and $1:n$ relationships between workflow steps are trivial to implement, e.g., one function sends multiple HTTP requests to other functions, creating a join pattern, i.e., $n:1$, is more challenging.
Current approaches involve either creating an additional function that is responsible for the join step or communicating through databases~\cite{barcelona-pons_faas_2019-1}.
In the first case, the function spends most of the invocation in an idle state, which leads to double-billing, increasing cost~\cite{baldini_serverless_2017}. 
In the second case, additional data storage is required, which was not intended to be used for communication between functions and also increases cost~\cite{elshamy_study_2023}.
In addition to higher cost, building such complex orchestration mechanisms defeats the purpose of the high levels of abstractions in FaaS platforms.

\section{Multi-Event Triggers}\label{sec:arch1}

With current serverless workflows, application developers still need to know a lot about how the platforms work internally to understand what kind of workloads are even possible to implement.
As this is against the spirit of serverless, we propose to make it easier to write workflows by introducing multi-event triggers.
These triggers make it trivial for application developers to implement a fan-in or join mechanism by specifying a \emph{set} of events that need to be collected before a function is triggered.

While different kinds of multi-event triggers are conceivable, we focus on event counts, \emph{ANDs}, and \emph{ORs} for multi-event trigger rules.
A necessary precondition for such rules is the existence of different event types.
In general, a multi-event trigger rule is specified by one or more event types and their respective quantity.
An event type can be, for example, the kind of sensor or client device that produces the event, or a more granular application-specific event type.

\begin{lstlisting}[float, label=lst:trigger_conditions_formal, caption={A formal specification of the structure of multi-event trigger rules.}, captionpos=b]
<rule>      ::= <count> ":" <type> |
    <condition> "(" <rule> "," <rule> ")"
<condition> ::= "AND" | "OR"
<count>     ::= "regexp:[0-9]+"
<type>      ::= "regexp:[a-zA-Z]+"
\end{lstlisting}

A multi-event trigger rule is made up of \emph{AND} and \emph{OR} conditions, which each contain two \emph{amount:event~type} pairs or, recursively, another rule.
A formal description is shown in~\cref{lst:trigger_conditions_formal}.

The simplest form of such a rule is a simple count-based trigger, where every n\textsuperscript{th} event of type $t$ results in a function call.
Consider the following example: Temperature readings (events) are produced periodically by a sensor, e.g., once per minute.
Once per hour, we want to trigger a function that uses a single measurement to store and analyze.
Hence, the trigger is defined by the triple of event type \emph{temperature}, the interval \emph{60 min}, and the function's URL.

More complex multi-event trigger rules combine different rules.
Consider the following smart home example.
Assume there are three sensors in an apartment: a temperature sensor, a wind sensor, and a motion sensor that produces an event when someone enters the apartment.
The wind and temperature sensors produce readings every 10 minutes.
There is a function that checks—depending on the value of its inputs—whether the heater should be turned on or off.
A trigger rule that implements this application is shown in \cref{lst:condition_example}.
In this example, the trigger rule is fulfilled once per hour (as both the wind and temperature sensors will each have produced six events) \emph{or} once there is a temperature reading and the motion sensor was triggered.
Once one of the two conditions is fulfilled, the function is invoked with the set of events that caused the condition to be fulfilled.
The function can use them to determine whether to turn on the heater.

\begin{lstlisting}[float, label=lst:condition_example, caption={An example of a trigger rule for a merge-based MET in a smart home setting.
    The trigger rule is fulfilled if either there a six temperature and wind measurements each or if there is one temperature measurement and a motion sensor detects a person entering the apartment.}, captionpos=b]
OR(
    AND(6:temperature,6:wind),
    AND(1:temperature,1:motion)
)
\end{lstlisting}

Internally, each trigger has one \emph{trigger set} for each event-type that is part of its \emph{trigger rule}.
These sets grow as events arrive, and—once the trigger rule is fulfilled—events are removed from the trigger sets and sent to the function.

There are limits as to which conditions are possible.
For instance, \emph{NOT} conditions are not possible, as we cannot guarantee while triggering a function that an instance of a certain event type has not been received.
This problem is exaggerated in the case of distributed, FaaS-based applications, in the case of network partitions or due to communication delays.

\section{The MET Engine}\label{sec:arch2}

\begin{figure*}
    \centering
    \includegraphics[width=0.66\linewidth]{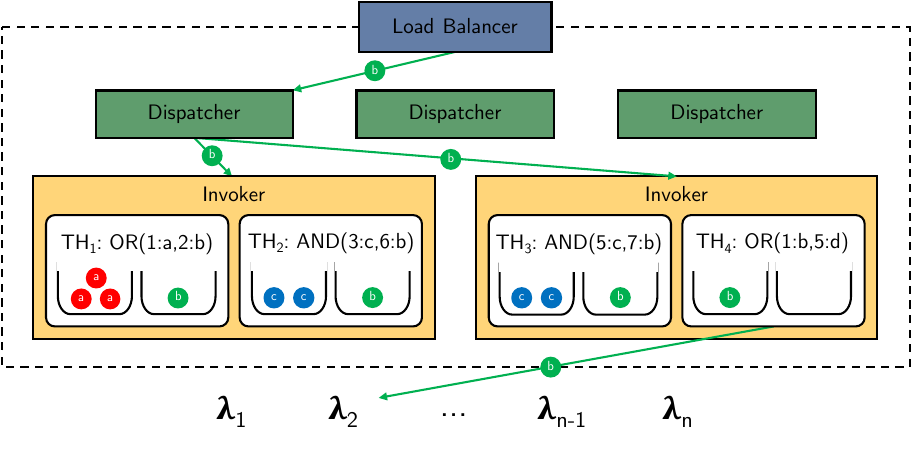}
    \caption{
        An overview of the MET engine's architecture.
        Different client devices produce events and send them to the MET engine.
        A load balancer distributes events between dispatchers, which process the events by forwarding them to the invokers that handle triggers using that event type.
        Each invoker handles a set of triggers, stores them, checks whether the trigger condition is fulfilled, and triggers the functions accordingly.
    }
    \label{fig:arch_overview}
\end{figure*}

We present an architecture for native platform support for multi-event triggers, the MET (multi-event trigger) engine.
The MET engine is a component that can be placed in front of existing FaaS platforms or that platforms can use to extend their function trigger mechanism, inhabiting a similar place to AWS's API Gateway, for example.
This architecture and our prototype (\cref{sec:implementation}) are designed for a multi-node, single rack setting, although the principles can be adapted to distributed or edge settings.

\cref{fig:arch_overview} shows an overview of the MET engine's architecture.
The system comprises two main components that can be scaled out individually: the dispatcher and the invoker.
In addition, a load balancer serves as central entry point to the system and distributes events between dispatchers.

The dispatcher receives events from the load balancer and forwards them to the appropriate invoker, which handles trigger logic.
For each MET, a new \emph{trigger handler} is created.
The trigger-handler contains a \emph{trigger set} for each event-type in which events are stored until the function is invoked.
For each event a trigger handler receives, it checks whether a trigger rule is fulfilled and, if that is the case, pulls the events from the trigger sets and invokes the function accordingly.

Dispatchers and invokers communicate using brokerless pub/sub messaging.
An invoker subscribes to events at the dispatchers based on the event types it has trigger rules for.
Consequently, the component handling the trigger logic (invoker) does not need to be concerned with handling events of types that do not match any of its triggers.
The dispatcher can be scaled out to different nodes (e.g., other VMs) in case load on the whole system increases to the point at which a single dispatcher cannot handle it alone.

Scalability is a key concern in designing MET engine.
The stateless nature of FaaS platforms allows distributed gateway designs, e.g., the frontend in AWS Lambda~\cite{brooker2023demand}.
Introducing state management, as in our trigger sets, complicates the design of MET engine.
We design the system so that it can handle the load realistically expected at a large FaaS platform.
As an example, consider AWS Lambda, which receives 10s of trillions of requests per month globally.
Assuming equal distribution across time and the 99 AZs, we can estimate the average load of individual AZs for Lambda to be in the high 10,000s or the low 100,000s requests per second\cite{barr2024lambda}.
We design the MET engine with a modular design as a distributed system to achieve these load parameters:
By deploying additional invokers, the amount of triggers that can be handled increases.
While our design currently imposes a single invoker as limit for a single trigger, we can circumvent this issue through \emph{trigger partitioning}.
By purposefully partitioning a MET into independent replicas, the amount of traffic it can handle increases even further.
This is made possible because these replicas do not need to communicate events, as long as the load-balancing dispatchers send an event to only one of the replicas.
While this impact the composition of event groups functions receive, this issue is negligible in cases where the order of incoming event only needs to be approximately kept.
Overall, the MET engine follows a \emph{multi-node, single rack} setting; meaning we distribute replicas of the dispatchers and invokers in the same datacenter with VMs.

\section{Prototype Implementation}\label{sec:implementation}

To demonstrate and evaluate our system, we implement an open-source proof-of-concept prototype.\!\footnote{\url{https://github.com/natalie-carl/MultiEventTriggers}}

\subsection{Context}\label{subsec:context}
In designing an architecture for handling METs at scale, we need to be able to handle the typical load large FaaS platforms experience.
In this context, we use multiple nodes in the same data center, as communication between servers in the same data center is efficient compared to settings where nodes are distributed geographically.
Simultaneously, this means that the maximum expected traffic for a single MET in our system is substantially less, requiring at most a single node.

\subsection{Infrastructure}
The dispatcher and invoker are written in Go.
We deploy them to a Kubernetes\footnote{\url{https://kubernetes.io/}} cluster, which we run on Google Kubernetes Engine.
Doing so allows us to easily adjust the number of dispatchers and invokers we want to use.
Each dispatcher and invoker runs in a separate pod.
The dispatchers send events to the invokers using ZeroMQ, a popular messaging library.\!\footnote{\url{https://zeromq.org/}}
Using a Kubernetes service of type \emph{LoadBalancer}, we distribute events between dispatchers.
The event producers publish their events by sending HTTP requests.
As individual triggers are registered at an invoker by providing a trigger rule and a function URL, our prototype can trigger arbitrary FaaS functions on every platform that supports HTTP.

\subsection{Trigger Rules}

Invokers receive trigger rules as specified in \cref{sec:arch1}, which they use to create a set of binary trees as internal representation of the MET.
It is noteworthy that an invoker needs to know which part of the trigger rules caused the rule to be fulfilled, as trigger rules might contain \emph{OR} conditions and invokers pull events from their trigger sets once the trigger rule is fulfilled.
In our prototype, we implement this by creating a binary tree for each case a trigger rule contains and checking them individually as a new event arrives.

\section{Evaluation}\label{sec:eval}

To evaluate our proof-of-concept prototype, we perform multiple experiments.
In the first experiment in \cref{subsec:exp1}, we measure the event–invocation latency compared to a baseline of manually handling event logic in the functions.
In the second and third experiment in \cref{subsec:exp2,subsec:exp3}, we measure the scalability of the MET engine when increasing concurrent requests and triggers, respectively.

\subsection{Experiment 1: Latency}\label{subsec:exp1}
To compare the overhead of using the MET engine versus handling events manually in a function and using an external database for keeping state, we implement a simplified data center incident detection application.

\paragraph{Use case}
We simulate three kinds of sensors that produce periodic events of types \texttt{temperature}, \texttt{packetLoss}, and \texttt{powerConsumption}.
We use a FaaS function to analyze the values these sensors produce and detect whether an incident occurred that requires manual intervention in the data center.
The trigger rule in \cref{lst:use_case_rule}  defines which combinations of events trigger the function.

\begin{lstlisting}[float, label=lst:use_case_rule, caption={Multi-event trigger rule used in our example use case.}, captionpos=b]
OR(
    AND(5:packetLoss,1:temperature),
    1:powerConsumption
)
\end{lstlisting}

The \texttt{packetLoss} and \texttt{powerConsumption} sensors produce single floating point numbers, while the \texttt{temperature} sensor produces an array of 25 floating point numbers to simulate different values for different racks in the data center.

\paragraph{Baseline}
To implement multi-event triggers in FaaS applications without a dedicated MET engine, we need a database for keeping state across invocations.
To this end, we use the PostgreSQL database system and deploy it to a virtual machine running in the same zone as the FaaS function to be invoked by the multi-event trigger.
The function invoked for each individual event and handles the event logic itself.
Consequently, most function invocations do not actually perform the incident detection, instead only resulting in a database transaction storing the new event.

\paragraph{Experiment setup}
We use the \texttt{k6} load generator\footnote{\url{https://k6.io/}} to simulate the sensors.
The \texttt{packetLoss}, \texttt{temperature}, and \texttt{powerConsumption} sensors use 20, 10, and 10 virtual users and produce 180, 36, and 18 events per minute, respectively.
The load generator runs for 30 minutes in each experiment run.
To compare the overhead different ways of handling complex event logic introduce, we are interested in \emph{event–invocation latency}: the duration between the creation of the event that causes a trigger rule to be fulfilled and the start of the execution of the application logic within the triggered function.
For the baseline, the function uses a timestamp created \emph{after} it handled the trigger logic but before the application part is run.
To ensure that we can compare timestamps from different sources, we deploy every virtual machine, Kubernetes cluster, and Cloud Run function in the same Google Cloud zone (\texttt{europe-west10-a}).

\begin{figure}
    \centering
    \includegraphics[width=\linewidth]{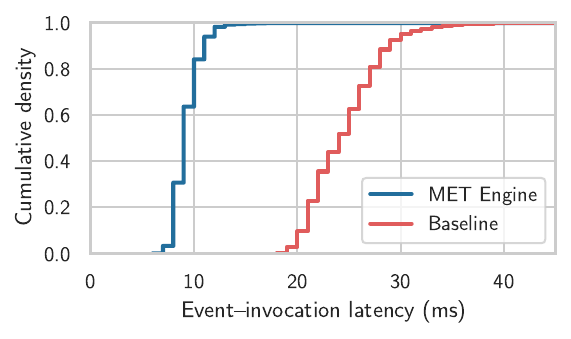}
    \caption{
        Cumulative distribution of latency values from the creation time of the last event that should trigger a function and the time at which the function logic actually starts.
        By using the MET engine over handling event logic within the function and coordinating database access across function instances, event–invocation latency is reduced by 62.5\% in the median.
    }
    \label{fig:exp1_results_ecdf}
\end{figure}

\begin{figure}
    \centering
    \includegraphics[width=\linewidth]{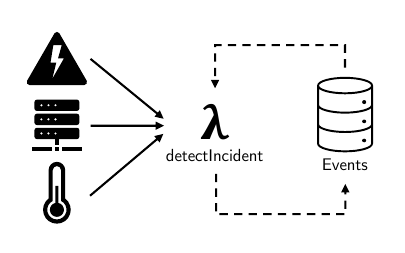}
    \caption{
        Baseline implementation for handling MET without the SUT.
        The load generator calls the function \texttt{detectIncident} for each event.
        The function handles the event logic in addition to the use case and uses a database for keeping state across executions.
    }
    \label{fig:exp1_baseline}
\end{figure}

\begin{figure*}
    \centering
    \includegraphics[width=0.85\linewidth]{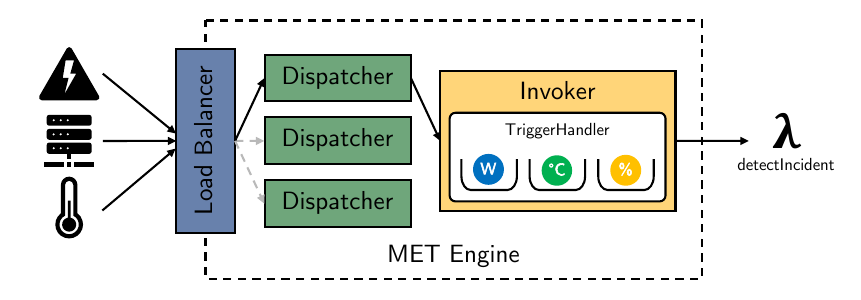}
    \caption{
        Experiment setup for measuring event–invocation latency with the SUT.
        The load generator sends requests to the MET engine's load balancer, which distributed events across three dispatchers.
        As we only use a single MET for the function, there is only one invoker.
        This invoker collects events until the trigger rule is fulfilled, after which the function \texttt{detectIncident} is called.
    }
    \label{fig:exp1_sut}
\end{figure*}

The system under test (SUT) is deployed to a Kubernetes cluster, which we run using Google Kubernetes Engine.
As we only use one MET in this experiment, we only use one invoker and, hence, only one node of type \texttt{e2-standard-4}.
The load balancer sends events to a load balancer, which forwards the request to one of three dispatchers.
The application—comprising the \texttt{detectIncident} function—runs on Google Cloud Run functions.
The baseline implementation is configured similarly, leaving out the Kubernetes cluster.
The function is modified to include the event logic, and we add a virtual machine in the same zone running PostgreSQL to store events.

\paragraph{Results}
\cref{fig:exp1_results_ecdf} shows the cumulative distribution of event–invocation latency values for both using the SUT and the baseline.
In the median, using the MET engine reduces the event logic's overhead by 62.5\%.
Furthermore, the function is called once for every event in the baseline case, which constitutes a $4.\overline{3}$ fold increase compared to using the MET engine.
This number depends on the trigger rule and the event arrival rates; it can only be calculated in cases where the event arrival rates are known.

\subsection{Experiment 2: Concurrent Requests}\label{subsec:exp2}

After demonstrating the event–invocation improvements, the following experiments focus on evaluating the performance of the MET engine.

\begin{figure}[!h]
    \centering
    \includegraphics[width=\linewidth]{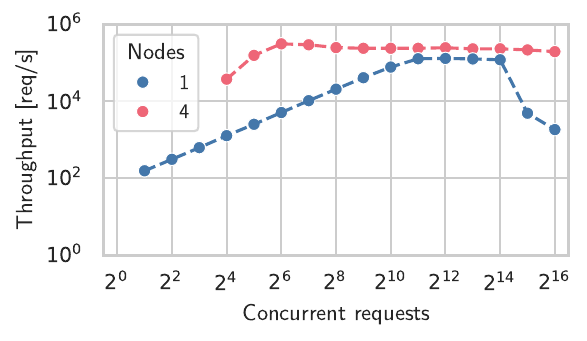}
    \caption{
        Throughput for different amounts of virtual users for a single-node and four-node cluster configuration.
        The single node reaches its maximum at 131,012.7 requests per second with 4,096 virtual users.
        With four nodes, maximum throughput increases to 313,154.81 requests per second.
    }
    \label{fig:exp2_throughput}
\end{figure}

\paragraph{Experiment description}

In the concurrent requests experiment, we evaluate how the MET engine scales with increasing amounts of concurrent requests with different cluster sizes.
To this end, we deploy the system under test as a single node and in a four-node configuration.
Each node is a \texttt{c7i.2xlarge} AWS EC2 virtual machine instance with 8 vCPUs and 16 GiB of memory, running in AWS's \emph{eu-central-1} region.
For load generation, we use a \texttt{c7i.16xlarge} EC2 instance with 64 vCPUs and 128 GiB of memory, running in the same region.
We create the trigger \texttt{(3:a)}, distributed across all nodes in the cluster.
As a result, a dummy function is invoked for every third event the trigger receives.
In addition, each request to the MET engine contains a 1,024-byte payload of randomly generated characters.
For both the configurations, we generate an increasing amount of concurrent requests.
For each amount of concurrent requests and cluster size, we generate load for five minutes and calculate the throughput as the average amount of requests processed per second by the system under test.
Between each such five-minute period, we leave sufficient time for the VMs to cool down.

\paragraph{Results}
\cref{fig:exp2_throughput} shows the results of the second experiment.
We observe that throughput increases linearly with the amount of concurrent requests up to $2^{11}$, at which point throughput is approximately 128,258.82 requests per second the single node scenario.
The maximum is reached at an average of 131,012.7 requests per second for $2^{12}$ concurrent requests.
Increasing the cluster size to 4 nodes increases maximum throughput to 313,154.81 requests per second, which is reached with $2^{6}$ concurrent requests.

\subsection{Experiment 3: Concurrent Triggers}\label{subsec:exp3}
The purpose of this experiment is to evaluate the effect of the amount of concurrent triggers on throughput within a single invoker.

\begin{figure}
    \centering
    \includegraphics[width=\linewidth]{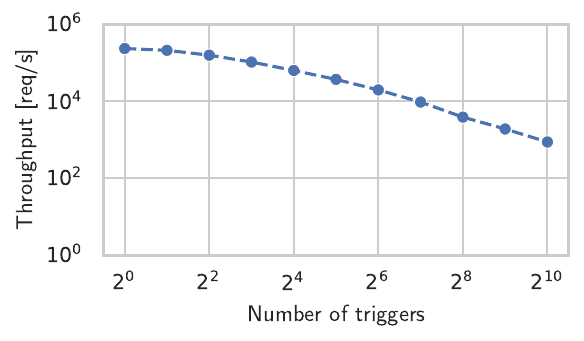}
    \caption{
        Throughput for increasing amount of concurrent triggers and a single invoker.
        While a single trigger with the trigger rule \texttt{AND(2:a,2:b)} reaches an average throughput of 236,601.77 requests per second, throughput decreases to 883.67 requests per second for 1,024 triggers on the same node.
    }
    \label{fig:exp3_throughput}
\end{figure}

\paragraph{Experiment description}
We run one invoker only on a \texttt{c7i.large} AWS EC2 virtual machine instance with 4 vCPUs and 8 GiB of memory.
For each step of the experiment, we run a constant number of copies of the same trigger for 5 minutes and measure the average throughput.
After each step, we have a cooldown period for resetting the system and increase the number of triggers exponentially.
To simulate having a variety of different triggers, we replicate the trigger \texttt{AND(2:a,2:b)} up to 1,024 times, each triggering a dummy function.
For each configuration, we have 128 virtual users, split evenly between event types $a$ and $b$.
Each request contains an additional 1,024-byte payload of random numbers, similarly to the previous experiment.

\paragraph{Results}
As expected, throughput decreases with increasing amounts of triggers running on the invoker.
For a single trigger, the invoker achieves an average throughput of 236,601.77 requests per second.
Increasing this to $2^{3}$ and $2^4$ results in 105,353.51 and 63,717.27 requests per second, respectively.
At the end, the invoker was only able to handle 883.67 requests per second.
After this point, the scheduling overhead was too large, and the system crashed.
The amount of concurrent triggers on a single invoker is primarily CPU-bound.
As more complex trigger rules require more computation for checking whether they are fulfilled, the invoker parallelized this process.

\section{Discussion and Future Work}

\subsection{Platform Integration}
Ideally, the MET engine would be tightly integrated into the FaaS platform itself.
This enables integration into the other features of the platform, such as the component that load balances incoming function calls.
We focus on a general implementation for our prototype of the MET engine instead, as our aim is demonstrating it viability for different use cases (and platforms).

\subsection{Stream Processing}
Stream processing and complex event processing are adjacent to multi-event triggers for FaaS.
Taking a broader view, both are concerned with determining when, \emph{and how often}, user-provided code is executed in response to incoming events.
With FaaS platforms that do not use the MET engine, \emph{every} event leads to an invocation.
Stream processing platforms, however, use windowing techniques to group events, for which a user-defined operator is then executed.
These engines do not include many features that developers depend on for serverless applications, such as scaling to zero, quickly adapting to bursty workloads, and support for diverse runtimes.
Thus, the MET engine approaches a similar set of problems from another direction.

\subsection{Geo Distribution}
The current implementation of the MET engine focuses on a multi-node, single rack setting.
Our evaluation shows that this setup scales enough to handle the load that AWS Lambda handles in a single availability zone.
A limitation of these design decisions is that it currently does not work well for geographically distributed multi-event triggers, which require tracking trigger state over more than a single rack.
In this case, the pub/sub component between invokers and dispatchers additionally needs to deal with network partitions and state drift between components.
We plan to explore this in future work, using conflict-free replicated data types (CRDTs) to track incoming events.

\subsection{Triggers}
Multi-event triggers present a new challenge for application developers on how to design effective trigger rules.
While the two triggers presented in this work (\texttt{OR} and \texttt{AND}) already enable complex use cases, we envision that application developers can also benefit from other types, such as \texttt{XOR}.

Another challenge is how to deal with synchronization and consistency in case of downtimes.
For example, using the trigger rule from \cref{lst:condition_example} where six temperature and six wind measurements were required, what happens if the wind sensor skips sending one event?
Currently, the MET engine would trigger the function with out-of-date temperature events (as it has to wait for an additional wind sensor event).
One possible solution to this problem is giving events a time-to-live, after which they can not trigger a function anymore.

\section{Related Work}\label{sec:relatedWork}

In most FaaS platforms, function invocations are stateless and correspond to single events.
Consequently, triggering a function in response to a collection of events only requires maintaining state to determine when a trigger condition is met~\cite{pfandzelter_iot_2019}.
Several approaches have introduced state management to FaaS.
\emph{Crucial}~\cite{barcelona-pons_faas_2019} adds a distributed shared object layer, essentially treating function instances as threads at larger scale and persisting state via an external server and S3 storage.
\emph{Cloudburst}~\cite{sreekanti_cloudburst_2020} integrates an autoscaling key-value store, co-locating data and functions to improve performance over public object storage.
Other systems employ alternative strategies for state management: \emph{Boki}~\cite{jia_boki_2021} introduces a serverless runtime that enables functions to persist and share state via a log-based API.
\emph{Faasm}~\cite{shillaker_faasm_nodate} leverages a WebAssembly runtime where function instances (\emph{Faaslets}) can share memory regions.
\emph{Photons}~\cite{dukic_photons_2020} achieves state sharing by co-locating function instances within the same execution environment.
\emph{Enoki}~\cite{pfandzelter2023enoki} combines an edge-native FaaS platform with geo-distributed data management for local data access for every function.
These stateful FaaS solutions could be adapted for multi-event triggers by maintaining state between function instances.

Our approach similarly uses an external component to keep state and store events that have not yet contributed to a function invocation.
However, unlike stateful FaaS systems, which nonetheless invoke functions for every incoming event, our system invokes functions only when a complete trigger condition is met, reducing the amount of unnecessary executions.
Additionally, using stateful FaaS would require a locking mechanism to prevent race conditions when function instances communicate via shared storage.
Our design avoids this issue by avoiding redundant invocations and handling event logic in the MET engine as part of the trigger mechanism itself.

FaaS function triggers have found further interest in the literature.
Scheuner et al.~\cite{scheuner_triggerbench_2022} benchmark FaaS triggers across multiple cloud platforms, finding that HTTP-based triggers outperform storage-based triggers, which suffer from higher latency.
This is additionally supported by Quinn et al.~\cite{quinn_implications_2021}, who observe significant delays when using object storage for workflow orchestration.
These findings further highlight the need for alternative state management and communication methods between function instances, especially for creating complex condition for when the application logic should be executed.

Function triggers serve as a central communication mechanism in serverless environments~\cite{scheuner_triggerbench_2022}.
Since serverless functions lack direct communication capabilities~\cite{hellerstein_serverless_2019}, triggers play a crucial role in orchestrating workflows.
As outlined in \cref{sec:background}, workflows are either managed by a centralized orchestrator or implemented as decentralized function choreographies, resulting in different workflow structures and execution patterns.

Taibi et al.~\cite{taibi_patterns_2020} describe the fan-in/fan-out pattern, also known as fork/join, where a task is divided into parallel executions across multiple function instances.
While most public FaaS platforms provide workflow engines (e.g., AWS Step Functions and Google Cloud Workflows), these solutions often exhibit inefficiencies for fork/join workflows~\cite{barcelona-pons_faas_2019-1}.
Among public platforms, IBM Composer was identified as the most effective for this use case~\cite{barcelona-pons_faas_2019-1}, but its deprecation alongside IBM Cloud Functions removed it as an option.

Several alternative approaches have been proposed, differing primarily in their scheduling mechanisms.
\emph{TriggerFlow}~\cite{lopez_triggerflow_2020} provides a framework for constructing custom workflow engines on Knative, where triggers are represented as user-defined Python callables, allowing for flexible trigger conditions.
Other systems distribute workflow scheduling across FaaS worker nodes, as seen in \emph{FaaSFlow}~\cite{li_faasflow_2022}.
\emph{DataFlower}~\cite{li_dataflower_2023} moves beyond control-flow-based orchestration by scheduling functions based on data availability, ensuring that functions execute only when all required inputs are ready, thereby reducing idle wait times.

In the domain of machine learning workflows, \emph{Cirrus}~\cite{carreira_cirrus_2019} employs a client-side backend for scheduling, while \emph{PyWren}~\cite{jonas_occupy_2017} coordinates fork/join steps using external storage and virtual machines for parallel data processing.
Alternatively, some approaches eliminate the need for a centralized orchestrator altogether, instead leveraging function-level libraries to enable choreography-based workflows~\cite{carl_geoff_2024,liu_doing_2023}.

Serverless computing has also gained traction in scientific workflows due to its potential for high parallelism and scalability.
Elshamy et al.~\cite{elshamy_study_2023} compare serverless and traditional serverful orchestration approaches, highlighting the necessity of additional storage layers in both models.
They identify two strategies for implementing a reduce stage: (1)~using a dedicated function that periodically checks for completion and (2)~embedding workflow logic within functions so that each instance verifies completion at the end of execution.
Similarly, Malawski et al.~\cite{malawski_serverless_2020} evaluate scientific workflows on public FaaS platforms and find comparable performance to serverful alternatives, further demonstrating the viability of serverless architectures in data-intensive applications.

A central problem serverless workflows face is that functions lack directly communication, which requires external storage or non-serverless components for coordination.
By extending the function trigger mechanism, we propose multi-event triggers as an alternative.

\section{Conclusion}\label{sec:conclusion}

In this paper, we introduced multi-event triggers as a novel extension to the event-based invocation mechanism in current FaaS platforms.
By enabling functions to be triggered based on complex combinations of multiple events—rather than individual occurrences—MET addresses a fundamental limitation in the current design of serverless workflows, namely the lack of native fan-in support and the inefficiencies associated with manually implementing such behavior at the application level.

To realize this concept, we proposed the MET engine, which can be integrated with existing FaaS platforms and decouples the evaluation of complex trigger conditions from the function logic itself.
Our evaluation demonstrates that this approach significantly reduces invocation latency—by 62.5\% in an incident detection use case, while also lowering resource consumption by reducing the number of superfluous function invocations.

Furthermore, our scalability experiments show that the MET engine can sustain high throughput under load, reaching over 300,000 requests per second in a four-node configuration.
These results indicate that the proposed architecture is suitable for deployment in large-scale cloud environments and can be integrated with minimal impact on existing systems.

In doing so, this work preserves core benefits of the serverless paradigm, such as loose coupling, automatic scalability, and minimal operational overhead, while enabling more expressive and efficient event-driven workflows.
Future work will focus on extending support for geographically distributed deployments, extending the expressiveness of trigger rule specifications, and exploring techniques for improved synchronization, fault tolerance, and consistency across distributed trigger handlers.

\newpage

\begin{acks}
    Funded by the \grantsponsor{DFG}{Deutsche Forschungsgemeinschaft (DFG, German Research Foundation)}{https://www.dfg.de/en/}—\grantnum{DFG}{495343202}.
\end{acks}

\balance

\bibliographystyle{ACM-Reference-Format}
\bibliography{met.bib}

\end{document}